\documentstyle[preprint,aps]{revtex}
\tighten
\begin{document}
\draft

\title{Quantum Fluctuations in the Equilibrium State of a Thin
Superconducting Loop}
\author{F.W.J. Hekking$^{(1,2)}$ and L. I. Glazman$^{(1)}$}
\address{$^{(1)}$Theoretical Physics Institute, University of Minnesota,
Minneapolis, MN 55455\\
$^{(2)}$Cavendish Laboratory, University of Cambridge, Madingley Road, Cambridge
CB3 0HE, United Kingdom\\}
\date{\today}
\maketitle

\begin{abstract}

We study the oscillatory flux dependence of the supercurrent in a thin
superconducting loop, closed by a Josephson junction. Quantum fluctuations of
the order parameter in the loop affect the shape and renormalize the amplitude
of the supercurrent oscillations. In a short loop, the amplitude of the
sinusoidal flux dependence is suppressed. In a large loop, the supercurrent
shows a saw-tooth dependence on flux in the classical limit. Quantum
fluctuations not only suppress the amplitude of the oscillations, but also smear
the cusps of the saw-tooth dependence. The oscillations approach a
sinusoidal form with increasing fluctuation strength. At any finite length
of the loop, the
renormalized current amplitude is finite. This amplitude shows a power-law
dependence on the junction conductance, with an exponent depending on the
low-frequency impedance of the loop.
\end{abstract}

\section{Introduction}
\label{introduction} Quantum effects in ultrasmall Josephson junctions have been
studied intensely now for more than a decade, both
experimentally\cite{devoret92} and theoretically\cite{schoen90}. The most
important manifestation of quantum fluctuations is the well-known macroscopic
quantum tunneling of the phase across a current-biased junction\cite{devoret92}.
This phenomenon leads to the observation of events of quantum phase slip at a
bias that is relatively close to the critical current.

More generally, macroscopic quantum tunneling causes a finite voltage to appear
at any finite current. The corresponding $I-V$ characteristic may be nonlinear,
and in the limit of zero temperature one finds $V=AI^\gamma$, with an exponent
$\gamma$ which depends on the impedance of the leads. This relation may be
derived in the framework of the Caldeira-Leggett model\cite{caldeira} that
treats quantum transitions between neighboring minima of a tilted washboard
potential in the presence of a dissipative ``environment''. It has been shown in
Refs.~\cite{weiss,fisher85a} that the coefficient $A$ is proportional to the
square of the ``bare'' (i.e., unaffected by the environment) tunnel matrix
element for transitions between the two minima of the potential. The dual result
for a voltage-biased junction shows that the DC current in such a junction is
proportional to the square of the Josephson energy of the junction, also
unrenormalized by the environment\cite{averin,falci,ingold94}.

Actually, these two complementary results follow from very similar treatments of
two closely related models, the Caldeira-Leggett model\cite{caldeira}, and the
electromagnetic environment model\cite{devoret90,girvin}: The effective boundary
conditions for the quantum fluctuations of the "environment modes" in these
treatments do not depend on the Josephson energy of the junction itself. This
approach (adequate for most existing experiments) is absolutely legitimate in
the case of weak fluctuations of the phase of the order parameter across the
junction. A more cautious analysis is needed, though, if these fluctuations are
strong. Indeed, quantum phase fluctuations in a one-dimensional (1D)
superconductor are known to diverge logarithmically with length. If these
fluctuations would result in a diverging random phase across the junction, the
Josephson energy of the system would average to zero, and neither of the quoted
results would be true.

We will show that a finite renormalized Josephson energy can arise because the
junction itself affects the fluctuations of the
environment~\cite{hekking}. Simultaneously, the
modes of the environment renormalize the plasmon oscillations in the junction.
This mutual influence of different parts of the circuit makes the separation of
it on two entities -- the junction with a fixed capacitance, and the environment
-- to be somewhat a matter of convention.

We consider a thin superconducting loop which contains a Josephson junction. The
"environment modes" of the loop consist of propagating plasmon modes with a
soundlike dispersion\cite{mooij}. The most straightforward way to observe a
possible renormalization of the Josephson energy, is to phase-bias the junction
by threading a flux $\Phi$ through the loop and measure the loop magnetization,
which is proportional to the Josephson current $J=J(\Phi)$. The renormalization
leads to values of the critical current
$J_c$ which are smaller than one would expect from the mean-field result,
$J_c^0=\pi \Delta G /(2e)$, where $G$ is the conductance of the junction, and
$\Delta$ is the superconducting gap in the loop. For a large loop, it also
changes the flux dependence $J(\Phi)$.

The paper is organized as follows. The model for the loop with the junction is
presented in Section~\ref{model}, followed by a qualitative discussion of the
dependence of $J$ on $\Phi$ in Section~\ref{qualitative}. The problem of the
renormalization of the Josephson coupling is treated in
Section~\ref{renormalization} where we make use of the similarity to the problem
of quantum Brownian motion in a periodic potential~\cite{fisher85b}. The effect
of the macroscopic quantum tunneling on the phase-dependence of the Josephson
current for a relatively large loop is discussed in Section~\ref{tunneling}. We
employ the analogy with the problem of pinning of a 1D charge density
wave\cite{larkin}, or Wigner crystal\cite{glazman}, and use an instanton
approach~\cite{vainshtein} to calculate $J(\Phi)$. Concluding remarks can
be found in Section~\ref{discussion}.

\section{The Model}
\label{model} We consider a superconducting wire of length $L$ and small
cross-sectional area $S=a\times a$, which is embedded in a medium with
dielectric constant $\varepsilon$. The wire is closed by a Josephson junction to
form a loop. The bare Josephson energy of the junction is $E_J^0 \equiv \pi
\Delta \hbar G/(4e^2)$, and its charging energy is $E_c = 4e^2/C$. Perpendicular
to the loop, a magnetic field ${\cal H}$ is applied, such that a flux ${\cal H}
L^{2}/(4\pi)$ threads the loop. We introduce the corresponding phase $\Phi =
{\cal H} L^{2}/(2 \Phi _{0})$, where $\Phi _{0}$ is the superconducting flux
quantum.

The low energy excitation spectrum of this system can be described in terms of
the phase of the order parameter $\varphi (x)$ by the Lagrangian ${\cal L}=K-U$,
where
\begin{equation}
K=\int_0^{L}dx\frac{\hbar^2}{2e_c}\left[\dot{\varphi}(x)\right]^2
+\frac{\hbar^2\left[\dot{\varphi} (L)-\dot{\varphi} (0)\right]^2}{2E_c},
\label{K}
\end{equation} and
\begin{eqnarray} U=\int_0^{L}dx\frac{\hbar^2n_sS}{8m}
\left(\frac{\partial\varphi(x)}{\partial x}-\frac{\Phi}{L}\right)^2 \nonumber \\
- E_J^0\cos[\varphi (L)-\varphi (0)].
\label{U}
\end{eqnarray} Here, $1/e_c=[8 e^2\ln(R/a)/\varepsilon]^{-1}$ is the
characteristic inverse charging energy per unit length of the loop ($R$ is the
distance to a metallic screen\cite{capacity}), $n_s$ is the density of the
superconducting condensate, and $m$ is the electron mass.

The first term on the right hand side of Eq.~(\ref{K}) together with the first
term on the right hand side of Eq.~(\ref{U}) describe the propagating plasma
mode~\cite{mooij} along the loop~\cite{buisson}. They correspond to the
electrostatic energy stored in the plasmons and to the energy associated with
the supercurrent in the wire, respectively. In the latter energy, we included
only the kinetic inductance, assuming the wire is thin and electrodynamic
effects are weak. The plasma mode is characterized by a linear dispersion
relation $\omega (k) = v_{\rm pl} k$ between frequency $\omega (k)$ and wave
vector $k$, where the plasma velocity is given by
\begin{equation} v_{\rm pl} =
\sqrt{\frac{e_c n_s S}{4m}} = c \frac{a}{2\lambda _L}
\sqrt{\frac{2 \ln (R/a)}{\pi \varepsilon}} . \label{dispersion}
\end{equation} Here, $c$ is the speed of light and $\lambda _L =
\sqrt{mc^2/(4\pi n_s e^2)}$ is the London penetration depth of the wire. At
temperatures $T$ much smaller than the superconducting gap $\Delta$, the plasma
mode involves oscillations of the supercurrent only, and damping due to
thermally excited quasiparticles is negligible. Retardation effects were
neglected in the derivation of (\ref{K}) -- (\ref{dispersion}), i.e., we assume
that $v_{\rm pl} \ll c$; hence we require $a \ll \lambda _L$.

The remaining terms in (\ref{K}) and (\ref{U}) refer to the Josephson junction.
For simplicity, we will completely neglect the junction capacitance, $C=0$,
throughout this paper. This approximation corresponds to neglecting the
last term in comparison with the first one in (\ref{K}) for all the relevant
scales of the phase variation. The shortest scale in the time dependence of
$\varphi$ is $\hbar/\Delta$, and correspondingly the smallest part of the ring
involved in the phase fluctuation is $\hbar v_{\rm pl}/\Delta$. Therefore,
we may set
$C$ to zero if
\begin{equation}
\frac{E_{c}}{\Delta} \gg \frac{mv_{\rm pl}}{(\hbar n_{s} S)}.
\label{C}
\end{equation}

An important length scale in our model is determined by the length $L^*$ at
which the energy of supercurrents in the loop and the Josephson energy of the
junction are of the same order. It is given by
\begin{equation}L^*\equiv\frac{\hbar^2n_sS}{4mE_J^0} =
\frac{\hbar g v_{\rm pl}}{\pi E_J^0},
\label{Lstar}
\end{equation}
where $g$ is defined through
\begin{equation}
\frac{1}{g} = \frac{4}{\pi} \frac{mv_{\rm pl}}{\hbar n_s S}= 8
\frac{e^2}{\hbar c}\frac{\lambda _L}{a}\sqrt{\frac{2\ln{(R/a)}}{\pi
\varepsilon}} .
\label{g}
\end{equation}
Here $e^2/\hbar c\simeq 1/137$ is the fine structure constant. Physically,
$1/g$ is the dimensionless zero-frequency impedance of the superconducting
wire~\cite{impedance},
$$
Z(\omega = 0)
=
\frac{1}{g}
\frac{2 \pi \hbar}{(2e)^2} .
$$

In the absence of fluctuations, the phase varies linearly with the distance
along the loop; therefore, it is convenient to introduce a new variable, $\chi
(x)$, by the relation:
\begin{equation}
\varphi (x) =\varphi_0 \frac{x}{L} +\chi (x), \label{chi}
\end{equation}
where $\varphi_0$ is determined such that the energy $U$ has its
minimum at $\chi =0$. This condition can be written in a simple form:
\begin{equation}
\sin\varphi_0 +\frac{L^*}{L}\varphi_0=\frac{L^*}{L}\Phi. \label{phi}
\end{equation}
For later use, we rewrite the expression~(\ref{U}) for $U$ in terms of
$\chi$ as follows:
\begin{eqnarray} U&=&E_J^0 \{ -\cos [\chi(2\pi)-\chi(0)+\varphi_0]
-[\chi(2\pi)-\chi(0)] \sin\varphi_0 \nonumber \\
&+&\pi\frac{L^*}{L}\int_0^{2\pi}d\theta\left(\frac{\partial\chi}{\partial\theta}
\right)^2 +\frac{L}{2L^*}\sin^2\varphi_0 \} . \label{Uang}
\end{eqnarray}
Here a new (``angular'') coordinate $\theta =2\pi x/L$ has been introduced.

The DC Josephson effect at zero temperature is described fully by the
$\Phi$-dependence of the ground state energy, $E_{\rm gr}$, of the system under
consideration. Because only the term $U$ of the energy depends explicitly on
$\Phi$, the persistent current $J(\Phi) \equiv (2e/\hbar)\partial E_{\rm
gr}/\partial\Phi $ can be expressed in terms of the average $\langle U\rangle$
over the ground-state wave function:
\begin{equation} J(\Phi) =
\frac{2e}{\hbar}\left(1+\frac{L}{L^*}\cos\varphi_0\right)^{-1}
\langle\frac{\partial U}{\partial \varphi_0}\rangle. \label{J}
\end{equation}

\section{Qualitative analysis}
\label{qualitative}

The behavior of $J$ as a function $\Phi$ is very different in the two limiting
cases $L \ll L^*$ and $L \gg L^*$, which we will discuss qualitatively below.

\subsection{The case $L \ll L^*$}
\label{qualitativeshort} For a relatively short loop, $L\ll L^*$, the kinetic
energy of the supercurrent dominates over the Josephson energy, and the phase
difference $\varphi _0$ across the junction is completely determined by the flux
threading the loop, $\varphi _0 \simeq \Phi$, as can be seen from
Eq.~(\ref{phi}). Classically, i.e., in the absence of phase fluctuations
($\chi =0$), the dependence of the persistent current on
$\Phi$ is given by $J(\Phi) = J_c^0\sin\Phi$, with $J_c^0 = 2eE_J^0/\hbar $. We
will see below that this classical result holds in the limit $1/g \to 0$.

In the quantum case $\chi \ne 0$, we can neglect the effect of the Josephson
junction on phase fluctuations when calculating the average in (\ref{J}) for $ L
\ll L^{*}$. In other words,
for a short loop only the term with the integral in (\ref{Uang}), which
corresponds to the kinetic energy of the supercurrent along the loop, is
important. Substituting Eq.~(\ref{Uang}) into Eq.~(\ref{J}), and neglecting
terms ${\cal O} (L/L^{*})$, we find
\begin{equation} J(\Phi) = J_{c}^{0} \langle \cos (\chi (2\pi) -\chi (0))\rangle
\sin \Phi .
\label{Jsmall}
\end{equation} The evaluation of the average $\langle \cos (\chi (2\pi) -\chi
(0))\rangle$ does not differ in fact from the well-known calculation of the
Debye-Waller factor\cite{ziman}. We quantize the fluctuating field $\chi
(\theta)$ in the standard way as follows:
\begin{equation}
\chi(\theta) =
\chi _0 +
\frac{1}{\sqrt{g}}\sum \limits_{n=1}^{\infty} \frac{1}{\sqrt{n}}
\cos{\left(\frac{n\theta}{2}\right)}\left[a_n^{\dagger} + a_n \right] ,
\label{chibose}
\end{equation} where operators $a _n$ satisfy canonical commutation relations.
Using~(\ref{chibose}), it is straightforward to evaluate
$\langle\cos(\chi(2\pi)-\chi(0))\rangle$ where the average is taken with respect
to the quadratic Hamiltonian
$$ H =
\sum \limits _{n=1}^{\infty}
\frac{\hbar v_{\rm pl}n \pi}{L} \left[ a^{\dagger}_{n}a_n + \frac{1}{2} \right].
$$ At low temperatures and $L\gg\hbar v_{\rm pl}/\Delta$, we finally obtain
$J(\Phi) = J_c \sin \Phi$, with a renormalized critical current \begin{equation}
J_c = J_c^0\left(\frac{\hbar v_{\rm pl}}{L\Delta}\right)^{1/g}
\left[ \frac{\pi L k_BT/\hbar v_{\rm pl}} {\sinh {(\pi L k_BT/\hbar v_{\rm
pl})}}\right]^{1/g}. \label{J_c}
\end{equation} The energy $\Delta$ in (\ref{J_c}) appears as a high energy
cutoff for the plasmon waves; the result $(\ref{J_c})$ holds for temperatures
$k_BT \ll \Delta$. The classical
result $J_{c} \to J_{c}^{0}$ is recovered in the limit $1/g \to 0$.

At this point we would like to note that the Lagrangian defined by
Eqs.~(\ref{K}) and (\ref{U}) has been derived in the limit of small phase
fluctuations, $\langle(\nabla \chi )^2\rangle \ll 1/\xi ^2(0)$. This poses an
upper bound on the allowed values of $1/g$, \begin{equation} 1/g \ll
\frac{e^2}{\hbar v_F}(k_F a)^2\frac{\ln (R/a)}{\epsilon}. \label{gmin}
\end{equation} The right hand side of this inequality is proportional to the
number $(k_Fa)^2$ of quantum channels in the wire, and typically is large. We
will be interested in superconducting wires characterized by $g \sim 1$. For
such wires, (\ref{gmin}) is satisfied, and the local quantum fluctuations of
the current due to the propagating plasma mode~\cite{mooij} are smaller than
the critical current. We therefore neglect phase slip
events~\cite{phaseslip,zaikin} in the wire.

Let us estimate the exponent $1/g$ for an Al wire. If the wire is very dirty,
with a mean free path $l\sim1$ nm, we estimate the zero temperature coherence
length to be $\xi (0) \sim 40$ nm and the London penetration depth to be
$\lambda _L(0) \sim 500$ nm. Present-day technology enables one to fabricate
wires with a cross-sectional area $S=a^2\sim (50$ nm$)^2$. Such wires would be
characterized by an exponent $1/g\sim 1.2$. For cleaner wires with $l \sim a$,
the exponent can be expressed as $1/g = (16/a($nm$))^{3/2}$. We conclude that
typical values of the exponent should be in the range $1/g \alt 1$.

Quantum fluctuations suppress the maximum Josephson current below its mean-field
value $J_c^0$. This suppression depends on the loop length $L$; according to
(\ref{J_c}) $J_{c} \to 0$ when $L\to\infty$. As we will
discuss below, this is an artefact of the lowest order of perturbation theory,
where the effect of the junction on the fluctuations in the attached wire is
disregarded completely.

\subsection{The case $L \gg L^*$}
\label{qualitativelarge}
As we have seen above, if $L \ll
L^*$, the solution $\varphi _{0} (\Phi)$ of equation (\ref{phi}) that provides
the absolute minimum of energy varies continuously with $\Phi$. For a large loop
with $L \gg L^*$, the solution $\varphi_0(\Phi)$ has discontinuities:
\begin{eqnarray}
\varphi _0 \simeq \frac{L^*}{L} \Phi &\makebox[1cm]{if}& 0 \le \Phi < \pi ,
\nonumber
\\
\varphi _0 \simeq 2 \pi + \frac{L^*}{L} (\Phi - 2 \pi) &\makebox[1cm]{if}& \pi
\le \Phi <2 \pi . \label{discont}
\end{eqnarray}
Because the Josephson energy dominates over the kinetic energy of
the supercurrents, the phase $\varphi _0$ remains ``pinned'' to the minima
of the
cosine potential. Correspondingly, in the absence of fluctuations the
equilibrium persistent current $J(\Phi)$ has cusps, \begin{eqnarray} J(\Phi)
\simeq J_c^0 \frac{L^*}{L} \Phi &\makebox[1cm]{if}& 0 \le \Phi < \pi , \nonumber
\\ J(\Phi) \simeq J_c^0 \frac{L^*}{L} (\Phi - 2 \pi) &\makebox[1cm]{if}& \pi \le
\Phi <2 \pi .
\label{cusps}
\end{eqnarray} We expect quantum fluctuations (i) to renormalize the bare
Josephson energy and thus to suppress the slope of the saw-tooth dependence;
(ii) to smear the cusps at $ \Phi = (2n+1) \pi$, as quantum tunneling will
remove the degeneracy between pairs of states having the same values of energy
but different values of $\varphi_0$. This is shown schematically in
Fig.~\ref{sawtooth}.

However, due to the fact that the Josephson energy is not a weak perturbation if
$L > L^*$, we have to take its effect on the quantum fluctuations into account.
This is very similar to the problem of pinning of a 1D crystal\cite{glazman}.
The decrease in $\langle\cos[\chi(2\pi)-\chi(0)]\rangle$ with a growing length
of the loop $L$ should saturate when $L$ exceeds the characteristic length
$L^*$. The saturation occurs because the Josephson coupling pins the
low-frequency modes, thus preventing the logarithmic divergence of the phase
fluctuations at the junction.

\section{Renormalization of the Josephson energy} \label{renormalization} In
this Section we will analyze the renormalization of the Josephson energy by the
charge fluctuations in the framework of the renormalization group (RG) approach.
This approach will enable us to treat both cases $L<L^*$ and $L>L^*$ in a
unifying manner. We will restrict our analysis first to zero applied flux, $\Phi
=0$. It is convenient to perform a Wick rotation to imaginary time $\tau$ and to
consider the euclidean action ${\cal S}$ for the fluctuating field $\tilde{\chi}
(\theta,\tau)
\equiv \chi(\theta, \tau) - \chi (2\pi-\theta,\tau)$, which can be easily
obtained from the Lagrangian ${\cal L}$. Next one integrates out the
fluctuations in $\tilde{\chi} (\theta,\tau)$ away from the junction, and obtains
the effective action for the field $\tilde {\chi}(\theta=0)$, \begin{equation}
{\cal S} =
\frac{\hbar g}{4 \pi} \int \frac{d\omega}{2\pi} |\tilde{\chi }(\omega )| ^2
|\omega | -\int d\tau E_J^0 \cos {\tilde{\chi }(\theta=0, \tau)} . \label{S_eff}
\end{equation} This action can be studied by a standard perturbative RG
method~\cite{fisher85b}. We introduce a running cut-off energy $\mu$, and find a
flow equation\cite{flowequation} for the dimensionless Josephson coupling energy
${\bar E}_J \equiv E_J /\mu$:
\begin{equation}
\frac{d{\bar E}_J}{dl} = ( 1 - 1/g ) {\bar E}_J + {\cal O}({\bar E}_J^3),\quad
dl = -d\mu /\mu . \label{flow}
\end{equation} This equation describes how the Josephson coupling $E_J$ is
renormalized when high-energy degrees of freedom are integrated
out~\cite{footnote}. From Eq.~(\ref{flow}) it follows that,
upon decreasing $\mu$, the energy ${\bar E}_J$ flows to zero if $g <1$, whereas
${\bar E}_J$ increases if $g >1$~\cite{schmid}. Note that in the latter case
perturbation theory breaks down as soon as ${\bar E}_J \sim 1$.

In order to investigate these cases in more detail, we integrate
Eq.~(\ref{flow}) from the high energy cutoff $\mu _h = \Delta$ at which $E_J =
E_J^0$ down to a value $\mu _l = \hbar v_{\rm pl}/l_0$, characterized by some
length $l_0$. As a result, we find
\begin{equation} E_J (\mu _l) = E_J^0 \left( \frac{\hbar v_{\rm
pl}}{l_0\Delta}\right)^{1/g} . \label{flowsol}
\end{equation}

\paragraph*{The case $g <1$.} In this case, the result (\ref{flowsol}) remains
valid for $\mu _l \to 0$, i.e., for the largest possible values of $l_0$.
Putting $l_0 \sim L$ we thus recover our earlier result~(\ref{J_c}) at $T=0$. At
any finite length $L > \hbar v_{\rm pl} /\Delta$, the renormalized Josephson
energy $E_J$ is smaller than the plasmon level spacing of the loop, $\hbar
v_{\rm pl} /L$. Therefore the perturbative analysis of
Section~\ref{qualitativeshort} applies, and we find for the flux-dependent
Josephson current
\begin{equation}
J(\Phi) =
\frac{2e E_J^0}{\hbar} \left(\frac{\hbar v_{\rm pl}}{L\Delta} \right)^{1/g} \sin
\Phi .
\label{J_csmallg}
\end{equation}
We see that quantum fluctuations will completely suppress the
Josephson current as $L \to \infty$ if $g <1$.

\paragraph*{The case $g >1$.} In this case, the situation is quite different.
The RG procedure should be stopped when ${\bar E}_J \sim 1$, i.e., when the
cutoff energy $\mu$ reaches a value $\mu _l$ which satisfies the condition $\mu
_l = E_J (\mu _l)$. As a result $E_J$ is renormalized down to a value $E_J^{\rm
eff}$, and should be determined from the condition of self-consistency,
\begin{equation} E_J^{\rm eff} = E_J^0 \left( \frac{ E_J^{\rm
eff}}{\Delta}\right)^{1/g} , \end{equation} which yields
\begin{equation}
E_J^{\rm eff}
=
E_J^0 \left(\frac{ E_J^0}{\Delta}\right)^{1 / (g-1)} .
\label{E_J^eff}
\end{equation}
The value $E_J^{\rm eff}$ is reached for $l_0 \sim (L^*/g)
(\Delta/E_J^0)^{1/(g-1)}$. We thus conclude that if $g >1$, the
result~(\ref{J_c}) at $T=0$ holds as long as $L \lesssim L^*$; for larger values
of $L$, the decrease of the Josephson coupling slows down, and eventually $E_J$
saturates at the value $E_J^{\rm eff}$ given by Eq.~(\ref{E_J^eff}). Further
suppression of the Josephson energy is prevented by the fact that the modes
$\tilde {\chi} (\omega)$ at frequencies $\omega < E_J^{\rm eff}/\hbar$ are
pinned by the Josephson coupling, and hence cannot participate in the
renormalization.

When using $\Delta$ as an upper energy cut-off in the derivation of
Eqs.~(\ref{flowsol}) -- (\ref{E_J^eff}), we assumed the condition (\ref{C}) to
be satisfied. In fact, the above treatment remains valid, even if (\ref{C}) is
violated, but the weaker condition $L^*\gg C$, holds; in this case, $\Delta$
should be replaced with $E_c$ in (\ref{flowsol}) -- (\ref{E_J^eff}).

The result (\ref{cusps}) for the flux-dependent Josephson current in a loop with
$L>L^*$ remains valid for values of flux away from the cusp at $\Phi = \pi$; we
just should replace
$E_J^0$ by $E_J ^{\rm eff}$. The behavior of $J(\Phi)$ for $\Phi \sim \pi$ is
strongly affected by quantum tunneling, which we will study in the next
Section.

\section{Quantum tunneling of phase}
\label{tunneling}

As we have seen in Section~\ref{qualitativelarge}, in the classical limit the
flux dependence of the persistent current has cusps at $\Phi = (2n+1)\pi$ for a
large loop, $L
\gg L^*$. At these values of $\Phi$, a degeneracy occurs: Two states having a
phase difference at the junction given by $\varphi _0 = 2n \pi$ and $\varphi _0=
2(n+1)\pi$ respectively, have the same energy. This degeneracy may be lifted by
the quantum fluctuations of phase at $1/g\neq 0$. Tunneling between the two
macroscopic states characterized by different values of $\varphi_0$ induces a
shift $\delta E$ of the ground state energy of the system. As a result, the
cusps in the function
$J(\Phi)$ will be smeared. We will show below that the tunnel splitting $\delta
E \ll \hbar v_{\rm pl}/L$, i.e., it is smaller than the gap between the
degenerate ground state and the first excited state of the loop with the
junction. Thus, at zero temperature, we are dealing with an effective two-state
system, and the flux dependence of $J(\delta \Phi \equiv \Phi - \pi)$ near $\Phi
= \pi$ will be given by ~\cite{zener}
\begin{eqnarray} &&J(\delta \Phi)
\approx
\frac{2e E_J}{\hbar} \frac{L^*}{L} \delta \Phi \nonumber\\ &&\times \left\{1 -
\frac{\pi}{\sqrt{(L^*/L)^2 \delta \Phi^2 + (\delta E /(\pi E_J))^2}} \right\}.
\label{smearing}
\end{eqnarray}
In particular, we see that the smearing is characterized by a
width $\delta \Phi _s \sim (L/L^*)(\delta E/ E_J )$,
see Fig.~\ref{sawtooth}.

The tunnel splitting $\delta E$ is proportional to the amplitude $t$ for
tunneling through a barrier of height $E_J$. This tunneling involves a varying
phase field $\varphi(x)$ along the loop, and therefore occurs in a
multi-dimensional potential landscape. The dominant contribution to the
tunneling action grows logarithmically with the system size $L$. For large $L$
the amplitude
$t$ can thus be obtained within the WKB-approximation. However the
pre-exponential factor has to be retained, as the leading term of the
WKB-approximation yields a power-law, rather than an exponential, decay of $t$
with $L$. The calculation of $\delta E$ is therefore conveniently performed with
the use of instanton techniques~\cite{vainshtein} which generalize the
WKB-method to higher dimensions and enable one to evaluate the pre-exponential
factor directly.

We start our anaysis by introducing a new phase variable $\phi (x) \equiv
\varphi (x) - \varphi (L-x) + (2x/L - 1)\Phi$ (note that $\phi(L/2) = 0$). In
imaginary time, the Lagrangian for this phase field reads \begin{eqnarray} {\cal
L} =
\frac{\hbar g v_{\rm pl}}{4 \pi}
\int \limits _{0}^{L/2} dx
\left[\frac{1}{v_{\rm pl}^2}\left(\frac{\partial \phi}{\partial \tau}\right)^2 +
\left(\frac{\partial \phi}{\partial x}\right)^2 \right] \nonumber \\ + E_J \cos
\phi (0,\tau).
\label{euclidean}
\end{eqnarray} Here, we put $\Phi =\pi$, i.e., we consider the degeneracy point.
The Josephson coupling energy $E_J$ appearing in~(\ref{euclidean}) is assumed to
be renormalized by the high energy degrees of freedom of the loop. It is a
complicated problem to actually perform such a renormalization procedure, due to
the strong anharmonicity of the potential $U$ (Eq.~(\ref{U})) close to the
degeneracy point. For our subsequent treatment of the quantum tunneling process
the following qualitative description of the renormalization scheme will be
sufficient. We imagine to integrate out high-energy degrees of freedom in the
spirit of Section~\ref{renormalization}, starting from $\Delta$ down to a
cut-off $\mu _l$, which satisfies the inequality $E_J^{\rm eff} \ll \mu _l \ll
\Delta$. These degrees of freedom are so fast that they "follow" the tunneling
process and merely adiabatically renormalize the Josephson energy $E_J ^0$ down
to the value $E_J$, where $E_J^{\rm eff} < E_J < E_J^0$. As will be discussed
below, the tunneling process itself consists of a slow part and a fast part. The
former involves the remaining low energy degrees of freedom with energies up to
$E_J$, whereas the latter involves those with energies between $E_J$ and $\mu
_l$.

We will be interested in tunneling between the initial phase configuration $\phi
_i(x, -{\cal T}/2) = -(2 \pi/L)(x-L/2)$ at time $\tau = -{\cal T}/2$ and the
final configuration
$\phi _f (x,{\cal T}/2) = (2 \pi/L)(x-L/2)$ at time $\tau = {\cal T}/2$. Both
are classical configurations which minimize the energy $U$, see Eq.~(\ref{U}).
The tunneling amplitude is characterized by the matrix element \begin{equation}
\langle \phi _f |e^{-H{\cal T}/\hbar}| \phi _i\rangle = N\int {\cal D}\phi
e^{-S/\hbar} .
\label{matrixelement}
\end{equation} Here, $N$ is a normalization constant, $H$ the Hamiltonian, and
$S = \int _{-{\cal T}/2}^{{\cal T}/2} d\tau {\cal L}(\tau)$ the action. The
matrix element~(\ref{matrixelement}) can be used to determine the ground state
energy $E_{\rm gr}$, because it decays as $e^{-E_{\rm gr}{\cal T}/\hbar}$ for
${\cal T} \to \infty$.

The matrix element~(\ref{matrixelement}) will be evaluated in the so-called
dilute instanton gas approximation~\cite{vainshtein}. We first will construct a
single instanton (SI), i.e., a classical trajectory in the inverted potential
$-U$ between the configurations $\phi _i$ and $\phi _f$ that passes once, at a
time $\tau _c$, through the minimum of
$-U$. According to Refs.~\cite{larkin,glazman}, the tunneling process consists
of a fast and a slow part. The actual tunneling at the junction (i.e., the
passage through the minimum at $\tau = \tau _c$) happens within a short time
$\tau _0$, and involves a part of the phase field with a length $L_0 = v_{\rm
pl} \tau _0$. The length $L_0$ is determined by minimizing the total action of
the SI; as we will see below $L_0 \sim L^*$, in agreement
with~\cite{larkin,glazman}. The rest of the phase field makes the transition
$\phi _i \to \phi _f$ slowly. Therefore the SI consists of three steps, see
Fig.~\ref{instanton}:

(i) $-{\cal T}/2 < \tau < \tau _c - \tau _0/2$: slow adjustment of the phase
field away from the junction from the initial configuration $\phi _i$ to the
intermediate configuration $\phi (x > L_0) = 0$;

(ii) $\tau _c - \tau _0/2 < \tau < \tau _c + \tau _0/2$: fast tunneling at the
junction involving a part of the phase field with length $L_0$;

(iii) $\tau _c + \tau _0/2 < \tau <{\cal T}/2 $: slow adjustment of the phase
field away from the junction from the intermediate configuration to the final
configuration $\phi _f$.

The matrix element~(\ref{matrixelement}) for a SI can be written as a product of
two amplitudes, one corresponding to the slow and one corresponding to the fast
contribution.

We describe the {\em slow adjustment} by decomposing the phase field $\phi
(x,\tau)$
into modes on the loop,
\begin{equation}
\phi (x,\tau) =
\sum \limits _{n=1}^{n_{\rm max}} \phi _n(\tau) \sin {(2\pi n x/L)} .
\label{decomposition}
\end{equation} The upper cut-off $n_{\rm max} \sim L/L_0$ indicates that slow
adjustment involves the phase field away from the junction ($x > L_0$) only. For
the initial (final) state of each mode we have
$\phi _n(\tau = -(+){\cal T}/2) = - (+) 2/n$; the dynamics of the modes is
determined by the Lagrangian~(\ref{euclidean}) without the $\cos $-term. This is
a quadratic problem and the contribution to the matrix
element~(\ref{matrixelement}) can be calculated exactly. We find
\begin{equation}
\langle \phi _f |e^{-H{\cal T}/\hbar}| \phi _i\rangle ^{\rm SI}_{\rm slow} =
\prod \limits _{n=1}^{n_{\rm max}} \sqrt{\frac{m_s \omega _n}{\pi \hbar}}
e^{-\omega _n {\cal T}} e^{-g/n} ,
\end{equation} where $m_s = 2\hbar g L/ (\pi v_{\rm pl})$ is the "mass" of each
mode and $\omega _n = 2 \pi n v_{\rm pl}/L$ its frequency. The SI action for
slow adjustment of the phase is easily calculated to be
\begin{equation} S^{\rm SI}_{\rm slow} \simeq \hbar g \log (L/L_0) .
\end{equation}

The {\em fast tunneling} of the phase field close to the junction ($x<L_0$) can
be described by the following Ansatz~\cite{larkin}: \begin{equation}
\phi (x,\tau) =
\phi _0(\tau) [1-x/L_0] \makebox[1cm]{;} -\pi < \phi _0 < \pi. \end{equation}
Substituting this into~(\ref{euclidean}), we find the Lagrangian for $\phi
_0(\tau)$:
\begin{equation} {\cal L}_0 =
\frac{\hbar g v_{\rm pl}}{4 \pi}
\left[
\frac{L_0}{3v_{\rm pl}^2}\left(\frac{d\phi _0}{d\tau}\right)^2 +
\frac{1}{L_0} \phi _0 ^2 \right] + E_J \cos {(\phi _0)} .
\end{equation} This Lagrangian describes the "rigid" tunneling of a part of the
phase field with length $L_0$ in terms of the motion of a particle with "mass"
$m_f = \hbar g L_0/(6\pi v_{\rm pl})$ and "coordinate" $\phi _0$ in an inverted
double-well potential $V(\phi _0)=-\hbar g v_{\rm pl}\phi _0^2/(4\pi L_0) - E_J
\cos (\phi _0)$. The action corresponding to a single passage through the
minimum of $V$ at $\phi _0=0$ can be estimated to be~\cite{larkin}
\begin{equation} S^{\rm SI}_{\rm fast} \simeq E_J L_0/v_{\rm pl} + \alpha \hbar
g, \end{equation} where $\alpha$ is a constant of order unity. Minimizing the
total SI action $S^{\rm SI}_t = S^{\rm SI}_{\rm slow}+S^{\rm SI}_{\rm fast}$
with respect to $L_0$ we find
$L_0 = \hbar g v_{\rm pl}/E_J \sim L^*$. Following the standard treatment for
tunneling in a 1D double well potential outlined in Ref.~\cite{vainshtein} one
can easily estimate the fast SI contribution to~(\ref{matrixelement}),
\begin{eqnarray} &&\langle \phi _f |e^{-H{\cal T}/\hbar}| \phi _i\rangle ^{\rm
SI}_{\rm fast} \nonumber \\ &&\sim
\sqrt{g}\frac{v_{\rm pl}{\cal T}}{L^*} \exp\left\{{-\frac{v_{\rm pl}{\cal
T}}{2L^*}}\right\}
\exp \left\{-\frac{S^{\rm SI}_{\rm fast}}{\hbar }\right\}. \end{eqnarray}

The total matrix element~(\ref{matrixelement}) can now be calculated in the
dilute instanton approximation~\cite{vainshtein}. One sums over all
configurations of single instantons and anti-instantons that involve
transitions $\phi _i \to \phi _f$ and $\phi _f \to \phi _i$ respectively.
This is done under the assumption that SI's do not
overlap, which is justified as long as $S^{\rm SI}_t \gg \hbar$, i.e., for
$(L/L^*)^g\gg 1$.
As a result we find
\begin{eqnarray} &&\langle \phi _f |e^{-H{\cal T}/\hbar}| \phi _i\rangle
\nonumber \\ &&\sim
\sqrt{g}\exp\left\{{-\frac{v_{\rm pl}{\cal T}}{2L^*}}\right\} \prod \limits
_{n=1}^{n_{\rm max}}
\left\{\sqrt{\frac{m_s \omega _n}{\pi \hbar}} e^{-\omega _n {\cal T}} \right\}
\nonumber \\ &&\times
\left[
\exp\left\{\frac{v_{\rm pl}{\cal T}}{L^*}e^{-S^{\rm SI}_t /\hbar }\right\} -
\exp\left\{-\frac{v_{\rm pl}{\cal T}}{L^*}e^{-S^{\rm SI}_t /\hbar }\right\}
\right].
\label{fullmatrixelement}
\end{eqnarray}

{}From the behavior of~(\ref{fullmatrixelement}) for ${\cal T} \to
\infty$ we infer that there are two low-lying energy eigenstates with energies
$E_{\rm gr}^{\pm} = E_{\rm gr}^0 \pm \delta E$, where $E_{\rm gr}^0$ is an
irrelevant reference energy and
\begin{equation}
\delta E
\sim
\frac{\hbar v_{\rm pl}}{L^*} e^{-S^{\rm SI}_t } \sim
\frac{\hbar v_{\rm pl}}{L} \left(\frac{L^*}{L}\right)^{g-1} . \end{equation} We
see that indeed $\delta E \ll \hbar v_{\rm pl}/L$ for large $L$, consistent with
the assumption leading to Eq.~(\ref{smearing}).

\section{Discussion}
\label{discussion}
The interplay between disorder and Coulomb correlations strongly influences the
properties of low-dimensional superconductors. This is well-known for thin
films~\cite{finkelstein}: A transition from superconducting (S) to insulating
(I) behavior occurs upon decreasing the thickness of the film. Only recently,
developments in fabrication techniques made experimental studies on {\em in
situ} grown quasi-1D wires~\cite{sharifi,herzog} possible. These indicate that a
similar transition might occur in a superconducting wire upon decreasing its
cross-sectional area $S$. Correspondingly, one may expect that if the parameter
$g$ is smaller than a certain threshold value $g_{c}$, the wire should behave
as an insulator on length scales even shorter than the loop circumference $L$,
and the theory presented above ceases to be valid.

Recent attempts to extend the
description of $T_c-$suppression in homogeneous thin films~\cite{finkelstein} to
include quasi-1D homogeneous wires have met with considerable
difficulties~\cite{eckern}. On the other hand, in a 1D boson system disorder is
known to induce a localized-delocalized transition, which occurs for strongly
attractive interactions between the bosons~\cite{giamarchi}. More specifically,
in terms of our model, the results of~\cite{giamarchi} would correspond to a
transition to insulating behavior at a value $g_{c}= 3/2$. This threshold in
interaction strength is reduced in the case of two coupled
chains~\cite{orignac}, and one may conjecture a reduction to $g_{c}=1$ for a
multi-mode wire, in agreement with~\cite{zaikin}.

An interesting model system which shows a S-I transition is a 1D array of
Josephson junctions~\cite{falci95,haviland}. The behavior of such an array is
determined by a competition of the Josephson coupling $E_j$ between the islands
(which favors a phase-coherent superconducting state) with the electrostatic
energy $E_0$ (which localizes Cooper pairs on the superconducting islands). If
the array is superconducting, $(E_j/8E_0)^{1/2} > 2\pi$, its low lying
excitations are 'phase waves' with a linear dispersion,
$\omega (k) \sim \sqrt{8E_j E_0} k$. We therefore speculate that a 1D array
containing a junction which is weakly coupled to its neighbors could be used to
study the renormalization of the Josephson energy discussed in this paper. An
advantage of this system is that the energies $E_j$ and $E_0$, which depend on
properties of the array, are well known and controllable in a typical
experiment~\cite{vdzant}; in particular the Josephson coupling $E_j$ can be
chosen from a large range of values. This also would enable one to
systematically probe the regime close to the S-I transition.

In conclusion, we considered a thin superconducting loop which contains a
Josephson junction. The "environment modes" of the loop, which consist of
propagating plasmon modes with a soundlike dispersion, were found to
renormalize
the Josephson energy of the junction. The strength of the
renormalization is determined by the dimensionless zero-frequency impedance
of the loop.
In order to observe this renormalization
we propose to phase-bias the junction by threading the loop with a flux and
measure the corresponding Josephson current. For a relatively short loop, the
kinetic energy of the supercurrent dominates over the Josephson energy and the
phase difference across the junction is completely determined by the flux
threading the loop. The supercurrent depends on flux in a sinusoidal fashion.
Quantum fluctuations suppress the amplitude of this dependence. In the opposite
limit of a large loop, the phase difference remains more or less "pinned" to the
minima of the Josephson energy. Correspondingly the persistent current shows a
saw-tooth dependence on flux in the classical limit. Quantum fluctuations not
only suppress the amplitude of the oscillations, but also affect
their shape. While the impedance of the loop increases, the cusps of the
saw-tooth dependence are smeared; as the impedance tends to the quantum unit
value the shape of the oscillations approaches a sinusoidal form.
At any finite length of the loop, the renormalized current amplitude is finite
and shows a power-law dependence on the junction conductance, with an exponent
depending on the impedance of the loop.

\acknowledgments
It is a pleasure to thank I. Aleiner, Y. Blanter, G. Falci, R. Fazio, A.
Finkel'stein, U. Eckern, D. Esteve, P. Joyez, C. Kane, A. Larkin, A. Schmid, G.
Sch\"on, U. Weiss, and A. Zaikin for useful discussions. L.G. acknowledges the
hospitality of the Max-Planck-Institute for Solid State Physics at Stuttgart and
Stuttgart University where this work was initiated. We furthermore acknowledge
the financial support of the Netherlands Organization for Scientific Research
(NWO), the European Community through contract ERB-CHBI-CT941764, and NSF Grant
DMR-9423244.

\begin{figure}
\caption{Schematic dependence of the Josephson current $J$ on
$\Phi$ for a large loop, $L \gg L^*$: (a) sawtooth dependence, found
in the absence of fluctuations; (b) quantum fluctuations suppress the slope
of the sawtooth dependence and smear the cusps over a typical width
$\delta \Phi _s$.}
\label{sawtooth}
\end{figure}

\begin{figure}
\caption{Four configurations of the phase field $\phi$ which occur during the
tunneling process; these configurations are discussed in the text.}
\label{instanton}
\end{figure}


\begin{references}
\bibitem{flowequation}This RG flow equation has been obtained and discussed
before in the literature, see Refs.~\cite{fisher85b,kane}.
\bibitem{zener}We will neglect Zener tunneling~\cite{ziman} across the
gap $\delta E$. This is correct as long as $\Phi$ changes
slowly with time, $\dot{\Phi} \ll L \delta E^{2}/(\hbar L^{*} E_{J})$.
\end{references}
\end{document}